\documentclass[sigconf]{aamas} 
\usepackage{balance} % for balancing columns on the final page

\setcopyright{ifaamas}
\acmConference[AAMAS '24]{Proc.\@ of the 23rd International Conference
on Autonomous Agents and Multiagent Systems (AAMAS 2024)}{May 6 -- 10, 2024}
{Auckland, New Zealand}{N.~Alechina, V.~Dignum, M.~Dastani, J.S.~Sichman (eds.)}
\copyrightyear{2024}
\acmYear{2024}
\acmDOI{}
\acmPrice{}
\acmISBN{}

\acmSubmissionID{<<EasyChair submission id>>}

\title[AAMAS-2024 Formatting Instructions]{Dual-Role AoI-based Incentive Mechanism for HD map Crowdsourcing}
\subtitle{Extended Abstract}
\author{Wentao Ye}
\affiliation{
  \institution{Chinese University of Hong Kong, Shenzhen\\ Shenzhen Institute of Artificial Intelligence and Robotics for Society}
  \city{Shenzhen}
  \country{China}}
\email{wentaoye@link.cuhk.edu.com}

\author{Bo Liu}
\affiliation{
  \institution{Shenzhen Institute of Artificial Intelligence and Robotics for Society}
  \city{Shenzhen}
  \country{China}}
\email{liubo@cuhk.edu.cn}

\author{Yuan Luo}
\affiliation{
  \institution{Chinese University of Hong Kong, Shenzhen \\
  Shenzhen Institute of Artificial Intelligence and Robotics for Society}
  \city{Shenzhen}
  \country{China}}
\email{luoyuan@cuhk.edu.cn}

\author{Jianwei Huang}
\affiliation{
  \institution{Chinese University of Hong Kong, Shenzhen\\
  Shenzhen Institute of Artificial Intelligence and Robotics for Society}
  \city{Shenzhen}
  \country{China}}
\email{jianweihuang@gmail.com}
\begin{abstract}
A high-quality fresh high-definition (HD) map is vital in enhancing transportation efficiency and safety in autonomous driving. Vehicle-based crowdsourcing offers a promising approach for updating HD maps. However, \rev{recruiting crowdsourcing vehicles involves making the challenging tradeoff between the HD map freshness and recruitment \pay}. \rev{Existing studies on HD map crowdsourcing often (1) prioritize maximizing spatial coverage, and (2) overlook the dual role of crowdsourcing vehicles in HD maps, as vehicles serve both as contributors and customers of HD maps.} \rev{This motivates us to propose the Dual-Role Age of Information (AoI) based Incentive Mechanism (DRAIM) to address these issues.} 
DRAIM aims to achieve the company's tradeoff between freshness and recruitment \pay.

\end{abstract}


\newcommand{\BibTeX}{\rm B\kern-.05em{\sc i\kern-.025em b}\kern-.08em\TeX}

\ifodd 1
\newcommand{\rev}[1]{{\color{black}#1}} %revise 
\newcommand{\com}[1]{\textbf{\color{red} (COMMENT: #1) }} %comment of the text
\newcommand{\comg}[1]{\textbf{\color{green} (COMMENT: #1)}}
\newcommand{\response}[1]{\textbf{\color{green} (RESPONSE: #1)}} %response to comment
\else
\newcommand{\rev}[1]{#1}

\newcommand{\com}[1]{}
\newcommand{\comg}[1]{}
\newcommand{\response}[1]{}
\fi
\def\dif{\mathop{}\hphantom{\mskip-\thinmuskip}\mathrm{d}}%
\let\daccent\d
\let\d\relax
\newcommand\d{\ifmmode\dif\else\expandafter\daccent\fi}

\def\pay{cost}
\def\reward{reward}

\makeatletter
\gdef\@copyrightpermission{
	\begin{minipage}{0.3\columnwidth}
		\href{https://creativecommons.org/licenses/by/4.0/}{\includegraphics[width=0.90\textwidth]{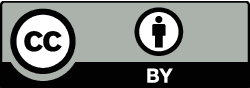}}
	\end{minipage}\hfill
	\begin{minipage}{0.7\columnwidth}
		\href{https://creativecommons.org/licenses/by/4.0/}{This work is licensed under a Creative Commons Attribution International 4.0 License.}
	\end{minipage}
	\vspace{5pt}
}
\makeatother

\begin{document}

%%% The following commands remove the headers in your paper. For final 
%%% papers, these will be inserted during the pagination process.

\pagestyle{fancy}
\fancyhead{}

%%% The next command prints the information defined in the preamble.

\maketitle 

%%%%%%%%%%%%%%%%%%%%%%%%%%%%%%%%%%%%%%%%%%%%%%%%%%%%%%%%%%%%%%%%%%%%%%%%

\section{Introduction}

A high-quality and fresh high-definition (HD) map is crucial for autonomous driving as it significantly enhances transportation efficiency and safety  \cite{xie2020energy, bao2022high, seif2016autonomous}. On the HD map, dynamic information, such as agglomerate fog, and the construction area, is vital in improving transportation efficiency \cite{dias2021cloud}. Moreover, the dynamic nature and unpredictability of this information necessitate its freshness, requiring updates from the HD map company (``company" hereafter) within seconds or minutes \cite{b2}. The conventional approach to updating HD maps involves the construction of dedicated mapping vehicle fleets equipped with high-precision sensors, such as LiDARs and high-accuracy cameras \cite{Ria2015build}. However, the high cost associated with maintaining such a fleet imposes limitations on its scale and the frequency of map updates. Therefore, vehicle-based crowdsourcing is a cost-effective approach to updating dynamic information on the HD map in academia \cite{cao2020trajectory, shi2023federated} and industry \cite{Mobileye2017Crowd,Here2021Mer}.

In practice, vehicles contributing to the HD maps can also benefit from utilizing them to enhance their driving experience. For instance, navigation, a fundamental feature of the HD map, is pivotal in improving transportation efficiency \cite{luo2019localization}. Therefore, vehicles involved in HD map crowdsourcing can serve as both contributors and customers to the HD map. This is one of the key differences between HD map crowdsourcing and general vehicle-based crowdsourcing, where vehicles only serve as contributors to perform the crowdsourcing tasks and receive the reward \cite{he2015high}.

Moreover, the recruitment \pay~ is a significant concern for HD maps. While crowdsourcing is often seen as a cost-effective means of data collection, maintaining an up-to-date HD map can be prohibitively expensive, particularly for frequent updates and extensive coverage. Since vehicles can serve as both contributors and customers to the HD map, HD map crowdsourcing allows the company to charge participating vehicles. However, previous works \cite{lai2019spir,cao2020trajectory}, neglecting this aspect, lead to the company overcompensating in pursuit of map freshness. To tackle this issue, we propose a flexible incentive scheme that allows for both positive and negative rewards, which constitute the company's recruitment \pay. A positive \reward~signifies compensation to vehicles, while a negative \reward~ denotes charges imposed on vehicles.

Alongside the consideration of recruitment \pay, the significance of ensuring freshness is emphasized in the realm of HD maps. However, many vehicle-based crowdsourcing literature focused on maximizing spatial coverage under the recruitment budget \cite{he2015high,hu2017duration}, which is unsuitable for HD map crowdsourcing. This is because the freshness is coupled with vehicles' trajectories. In other words, maximizing the spatial coverage is not enough to ensure the freshness of the HD map. Specifically, when recruiting crowdsourcing vehicles (``vehicles" hereafter) with the same spatial coverage, the resulting improvement in the HD map freshness can vary significantly. To characterize the freshness of the HD map, we use the expected age of information (AoI)  \cite{lou2022efficient, wang2022dynamic}. 

Another key characteristic of HD map crowdsourcing is that vehicles, acting as strategic players, strive to maximize their own payoff during their interaction with the company \cite{lai2019spir}. As customers of the HD map, vehicles obtain an HD map usage utility, which is influenced by the expected AoI of the HD map and their individual trajectories. On the one hand, vehicles benefit from a better HD map service with a smaller HD map AoI. For instance, an HD map with a smaller expected AoI can detect congestion more promptly and re-route vehicles to avoid it. On the other hand, vehicles only care about the freshness of locations along their specific trajectories. Therefore, despite using the same HD map, vehicles obtain varying HD map usage utility due to their distinct trajectories. To capture this characterization, our paper introduces the concept of trajectory AoI to help characterize the age- and trajectory-dependent HD map usage utility (``HD map usage utility'' hereafter).

Against this background, we design the \underline{D}ual-\underline{R}ole \underline{A}oI-based \underline{I}ncentive \underline{M}echanism (DRAIM) to incentivize vehicles considering the expected AoI of the HD map and regarding vehicles as both contributors and customers. Specifically, we model the interaction between the company and vehicles as a two-stage Stackelberg game. This model defines the company's payoff as the tradeoff between the HD map expected AoI and recruitment \pay~. On the other hand, vehicles' payoff encompasses their HD map usage utility, costs, and the \reward~received from the company. Moreover, our model accounts for vehicles' heterogeneous trajectories and costs.

\section{Model and Observations}
We consider a setting where the company wants to maintain an HD map with multiple kinds of dynamic information by recruiting vehicles. The interactions between the company and vehicles are as follows. The company \footnote{For convenience, we refer to the company as 'he' and the vehicle as 'she'.} offers HD map-based services, \emph{e.g.,} navigation\footnote{Notice that the HD map may provide multiple alternative trajectories to assist vehicles in navigating various accidental events and situations}, and monetary rewards to participating vehicles, which is determined by DRAIM. Note that the \reward~might be negative, which implies the company could charge for vehicles. In that case, the recruitment cost, the sum of rewards provided with vehicles, is also negative. Then, vehicles decide to participate or not. If she chooses to participate, she senses the traffic during her travel and sends sensing data to the HD map. After that, the company receives sensing data from those vehicles and updates the HD map. Simultaneously, the participated vehicle will receive the HD map services, \emph{e.g.,} navigation, and the \reward~from the company. We model this interaction as a two-stage Stackelberg game. In Stage I, the company decides on the reward to maximize her payoff. In Stage II, Each vehicle determines whether to participate or not to maximize his payoff. 

Our results give us the following observations:
\begin{enumerate}
    % \item If vehicles have similar marginal costs, which is defined as the difference between the cost and the HD map usage utility, the company has to pay for vehicles unless the vehicles' HD map usage utility of full participation, \emph{i.e.,} all vehicle participate, covers their cost. This is because if any vehicle type refuses to participate, it significantly reduces the HD map usage utility for all vehicles, resulting in a collective decision of non-participation. On the other hand, if the marginal costs of vehicles are far apart, the company must cover the costs unless at least one vehicle has a negative marginal cost of self-participation, \emph{i.e.,} only vehicles of the same type participate.
    \item If vehicles have similar marginal costs, which is defined as the difference between the cost and the HD map usage utility, and the HD map usage utility of full participation, \emph{i.e.,} all vehicle participate, can cover vehicles' costs, the optimal reward is non-increasing in the number of vehicles. This is because the company will occupy a more dominant position in the interaction as there are more vehicles. On the other hand, if the marginal costs of vehicles are far apart, the optimal reward may decrease in the number of vehicles.
    \item As more vehicles are involved, vehicles will eventually decline participation, even though more vehicles would result in a higher HD map usage utility. This is attributed to the company's decreased \reward. For vehicles, the detrimental effect of decreased \reward~ outweighs the potential benefit of a greater HD map usage utility for vehicles.
    \item The company may obtain a lower payoff as there are more vehicles if the freshest HD map usage utility of vehicles cannot cover their costs. This is because when vehicles' lowest marginal cost is positive, the optimal reward remains positive. However, the company has to pay for all participating vehicles, regardless of their number. In that case, the increased recruitment \pay~would gradually outweigh the benefit of the fresher HD map as the number of vehicles increases.
    % \item The company's payoff $U_c^*$ initially increases and then decreases as the vehicle type proportion $\gamma$ increases. This pattern emerges because when the proportion approaches $0.5$, the marginal costs of full participation $\A,\E$ reach their minimum. This signifies that vehicles are most willing to participate in HD map crowdsourcing at this moderate proportion. This finding suggests that the company should pay attention to the type proportion of crowdsourcing vehicles. By doing so, the company can harness the potential benefits and contributions from a broader diversity of vehicles, leading to a more cost-effective HD map crowdsourcing.
\end{enumerate}

\section{CONCLUSION AND FUTURE WORK}
This paper introduces DRAIM as a solution to encourage vehicle participation in HD map crowdsourcing. DRAIM leverages the dual role of vehicles to effectively minimize the company's costs associated with HD map updates, taking into account the tradeoff between freshness and recruitment costs.  %To solve it, we use a two-stage Stackelberg game to formulate the interaction between the company and vehicles. %Our findings surprisingly indicate that when vehicles' marginal costs of full participation are positive, increasing the number of vehicles may lead to a lower payoff for the company. %Moreover, the proposed DRAIM improves the company's payoff by $26.51\%$ on average, compared to the state-of-the-art mechanism that does not consider the dual role of vehicles.

In future research, we can explore scenarios where vehicles have more flexible options for interacting with HD maps. This includes studying situations where vehicles can utilize the HD map without actively contributing to it and cases where vehicles contribute to the HD map without utilizing it themselves.

\begin{acks}

This work is supported by the National Natural Science Foundation of China (Project 62271434, Project 62102343, and 62203309), Shenzhen Science and Technology Program (Project JCYJ20220818103006\\012, JCYJ20230807114300001, Grant No. JCYJ20210324120011032 and RCBS20221008093312031), Guangdong Basic and Applied Basic Research Foundation (Project 2021B1515120008), Shenzhen Key Lab of Crowd Intelligence Empowered Low-Carbon Energy Network (No. ZDSYS20220606100601002), and the Shenzhen Institute of Artificial Intelligence and Robotics for Society.
% If you wish to include any acknowledgments in your paper (e.g., to 
% people or funding agencies), please do so using the `\texttt{acks}' 
% environment. Note that the text of your acknowledgments will be omitted
% if you compile your document with the `\texttt{anonymous}' option.
\end{acks}

%%%%%%%%%%%%%%%%%%%%%%%%%%%%%%%%%%%%%%%%%%%%%%%%%%%%%%%%%%%%%%%%%%%%%%%%

%%% The next two lines define, first, the bibliography style to be 
%%% applied, and, second, the bibliography file to be used.

\bibliographystyle{ACM-Reference-Format} 
\bibliography{sample}

%%%%%%%%%%%%%%%%%%%%%%%%%%%%%%%%%%%%%%%%%%%%%%%%%%%%%%%%%%%%%%%%%%%%%%%%

\end{document}